\documentclass[prl,aps,showkeys,preprintnumbers,twocolumn,showpacs,amsmath,amssymb]{revtex4}
\usepackage{graphicx}
\usepackage{dcolumn}
\usepackage{bm}
\begin{document}
\preprint{\textit{Version February 11th 08}}
\title{Investigation of Non Resonant Dot - Cavity Coupling in Two Dimensional Photonic Crystal Nanocavities}
\author{M. Kaniber} \email{kaniber@wsi.tum.de} \homepage{http://www.wsi.tum.de}
\author{A. Laucht}
\author{A. Neumann}
\author{J. M. Villas-B\^oas}
\author{M. Bichler}
\author{M.-C. Amann}
\author{J. J. Finley}
\affiliation{Walter Schottky Institut, Technische Universit\"at M\"unchen, Am Coulombwall 3, D-85748 Garching, Germany}
\date{\today}
\begin{abstract}
We study the optical emission from single semiconductor quantum dots coupled to the optical modes of photonic crystal nanocavities. For dots that are both spectrally and spatially coupled, auto-correlation measurements reveal efficient single photon generation, with a drastically reduced lifetime due to the Purcell-effect. However, the multi-photon emission probability is enhanced compared to the same QD transition when it is detuned from the cavity mode by controlled $N_2$-deposition. This indicates the presence of an emission background that is shown to be related to the dot using photon cross-correlation spectroscopy. Photon temporal correlations persist even for large spectral detunings beyond $\Delta\lambda\sim-10~nm$, excluding the intrinsic QD continuum and phonon mediated processes as being responsible for the cavity mode emission background. We propose a mechanism based on photon induced shake up processes in the charged quantum dots, enhanced by the optical cavity.
\end{abstract}
\pacs{	73.21.La 
        42.70.Qs 
        42.50.Ct 
        42.50.Dv 
        85.60.Jb 
}
\keywords{Quantum dots, Photonic crystal, Single photons, Auto-correlation, Cross-correlation}
\maketitle
%
%
Single photon sources  based on semiconductor quantum dots \cite{Shields07,Michler00} (QDs) are needed for many applications ranging from quantum key distribution \cite{Gisin02} to quantum information processing using linear optical components \cite{Knill01}. In the latter case, the single photons should not only be emitted very efficiently \cite{Kaniber08} but also over timescales that are short compared with exciton dephasing processes \cite{Santori02}. This requirement can be fulfilled by incorporating QDs into solid state optical nanocavities and by using the Purcell effect to enhance efficiency and improve photon quantum indistinguishability \cite{Michler00,Pelton02,Chang06,Press07,Hennessy07}. Although the coupling of single quantum dots to the optical modes of such nanocavities has already been achieved by a number of groups worldwide \cite{Press07, Hennessy07,Moreau01, Hours03}, the majority of single dot experiments report a pronounced background emission in the vicinity of the cavity mode, even when a single QD transition is \emph{not} resonantly coupled. Recently, the emission from this background has been shown to be correlated with the emission from the detuned QD, indicating that a mechanism exists by which a spectrally detuned dot can emit into the cavity mode. Several mechanisms have been proposed to account for this emission background, including wetting layer-QD continuum transitions \cite{Hennessy07} and the presence of correlations has been suggested to be due to phonon assisted processes \cite{Press07}. However, no systematic investigations have appeared until now and conclusive statements about the origins of the background emission remain elusive.
%
%
\\
In this paper, we present comparative investigations of single photon generation from self-assembled In$_{0.5}$Ga$_{0.5}$As QDs in a photonic crystal (PC) defect nanocavity as a function of spectral detuning between the dot and the cavity mode. By performing auto- and cross-correlation spectroscopy, we examine the influence of the modified photonic environment on the spontaneous emission dynamics of individual dots and prove single photon generation. We find that QD transitions that are spectrally in resonance with the nanocavity mode exhibit a significantly enhanced multi-photon emission probability, due to background emission into the cavity mode \cite{Press07, Hennessy07}. By controlled adsorption of molecular Nitrogen into the PC nanostructure \cite{Mosor05} we detune the cavity mode, enabling us to systematically investigate the influence of the local photonic environment on both the optical properties of the same QD transition and the nature of this background emission. Photon cross-correlation measurements show that all emission lines investigated stem from the same dot. Most remarkably, pronounced cross-correlations are observed between the dot transitions and the cavity mode even for very large energy detunings ($>15~meV$). This surprising observation indicates the presence of an intrinsic coupling mechanism between the spectrally detuned QD and the cavity mode.\\
%
%
The samples studied are grown by molecular beam epitaxy and consist of the following layers grown on a semi-insulating GaAs wafer:  an undoped GaAs buffer followed by a 500 nm thick Al$_{0.8}$Ga$_{0.2}$As sacrificial layer.  This was followed by a $180~nm$ thick GaAs waveguide, at the midpoint of which a single layer of self-assembled In$_{0.5}$Ga$_{0.5}$As quantum dots was incorporated. A 2D-PC was formed by patterning a triangular array of cylindrical air holes using electron-beam lithography and reactive ion etching. The lattice constant of the PC was $a=280$~nm and the air hole radius $r=0.33a$.  Nanocavities were established by introducing three missing holes to form an $L3$ cavity \cite{Akahane03}. Finally, free standing GaAs membranes were formed by a $HF$ wet etching step. A scanning electron microscopy image of the investigated nanocavity is shown in the inset of Fig. 1a and an overview of the different fabrication steps can be found in ref. \cite{Kress05}.\\
%
%
The sample was mounted in a liquid He-flow cryostat and cooled down to $T=15~K$. For excitation we used either a pulsed Ti:Sapphire laser (f$_{laser}=80~MHz$, $2~ps$ duration pulses) or a continuous wave (cw) laser tuned into the wetting layer (WL) continuum at $\lambda_{exc}=837~nm$. The QD micro-photoluminescence ($\mu$-PL) was collected via a $100\times$ microscope objective ($NA=0.8$) providing a spatial resolution of $\sim700~nm$ and the signal was spectrally analyzed by a $0.55~m$ imaging monochromator and detected with a Si-based, liquid nitrogen cooled CCD detector. For time-resolved measurements we used a fast silicon avalanche photodiode that provided a temporal resolution of $\sim 100~ps$ after deconvolution. A pair of similar detectors in Hanbury Brown and Twiss configuration were used for both photon auto- and cross-correlation measurements.\\
%
\begin{figure}[ht]
    \begin{center}
        \includegraphics[width=\columnwidth]{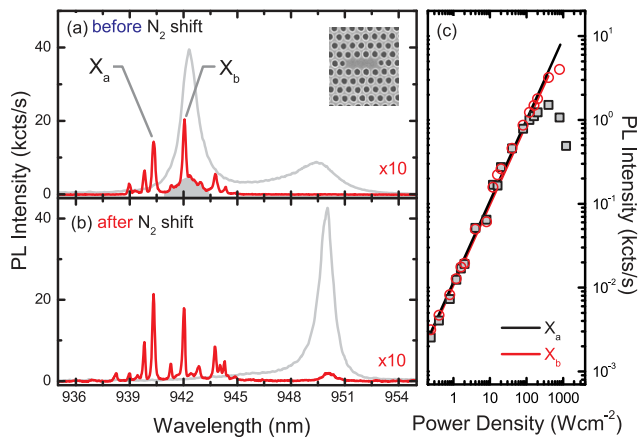}
    \end{center}
    \caption{ (color online) (a) $\mu$-PL spectra recorded from a single QD coupled to a cavity mode for weak (red) and strong (gray) 	
    	         cw pumping. The gray shaded region indicates the mode background emission. (inset) Scanning electron micrograph of the L3 cavity  
    	         investigated. (b) $\mu$-PL spectra of the same QD and the spectrally shifted cavity mode after N$_2$-deposition. (c) PL intensity as a  
    	         function of the excitation power for the $X_a$ and $X_b$ line. }
\end{figure}
%
%
In Fig. 1a we present $\mu$-PL measurements of a single QD that is spectrally and spatially coupled to a PC nanocavity. For weak cw excitation ($0.4~W/cm^2$) we observe several emission lines, all stemming from different transitions of the same QD (Fig. 1a - red trace). The two most prominent lines, labeled $X_a$ ($\lambda_{X_a}=940.33~nm$) and $X_b$ ($\lambda_{X_b}=942.04~nm$) in the figure, are \emph{out of}- and \emph{in}-resonance with the cavity mode ($\lambda_{cav, 0}=942.32~nm$). The cavity mode can be clearly observed at higher excitation powers ($400~W/cm^2$) and has a quality factor of 850 (Fig. 1a - gray trace). Both $X_a$ and $X_b$ show a linear power dependence (Fig. 1c) demonstrating that they arise from single exciton transitions. Cross-correlation measurements between $X_a$ and $X_b$ reveal that both emission lines arise from the same QD (presented below) and probably arise due to different charged states. When compared to $X_a$, transition $X_b$ saturates at higher excitation power, as can clearly be seen by examining Fig. 1c. This is a clear sign of Purcell enhanced emission due to the spectral proximity of cavity mode \cite{Kaniber07}, a hypothesis tested below using time resolved spectroscopy. To investigate the influence of the cavity on the emission, we spectrally shifted $\lambda_{cav}$ using a controlled $N_2$-deposition technique \cite{Mosor05}. A typical result is shown in Fig. 1b, where $\lambda_{cav}$ has been detuned by $\Delta\lambda=+7.7~nm$ to $\lambda_{cav, N_2}=950.2~nm$. This procedure allows us to investigate the \emph{same} QD state $X_b$ when it is spectrally coupled  to the cavity mode (\textit{before} N$_2$-deposition) and when strongly detuned (\textit{after} N$_2$-deposition). By comparing the (low power) $\mu$-PL spectra in Fig. 1a and 1b, we note that no emission is observed close to $\lambda_{cav, N_2}=950.2~nm$ before the $N_2$ shift was executed \cite{comment02}. This indicates the presence of a background emission associated with the cavity mode as was recently reported in refs. \cite{Press07, Hennessy07}.\\
%
\begin{figure}[ht]
    \begin{center}
        \includegraphics[width=\columnwidth]{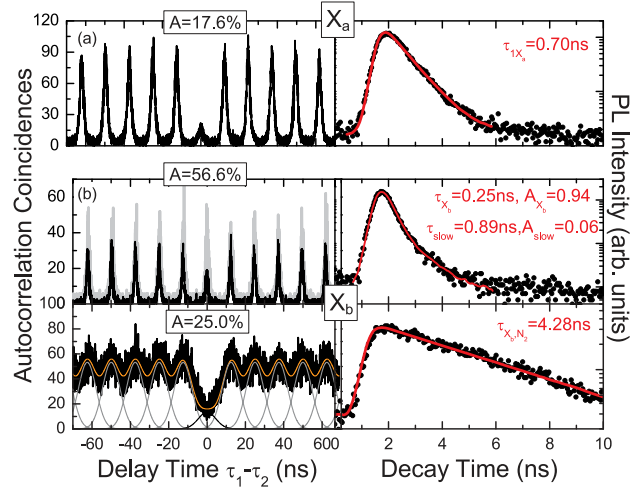}
    \end{center}
    \caption{ (color online) (Left panels) Auto-correlation measurements and corresponding time-resolved data (right panels) for (a) $X_a$ and (b) $X_b$. The two panels in (b) compare data for $X_b$ coupled to the cavity mode \textit{before} N$_2$-deposition with the case when $X_b$ is spectrally detuned from the cavity mode \textit{after} N$_2$-deposition. }
\end{figure}
%
%
To confirm that we probe transitions of a single QD, we performed photon auto-correlation and time-resolved spectroscopy measurements on $X_a$ and $X_b$ when subjected to pulsed optical excitation. These measurements were performed with a $\Delta\lambda=0.3~nm$ wide detection window centered on the transition of interest and the results are presented in Fig. 2a for transition $X_a$ and Fig. 2b for $X_b$. The upper and lower panels of Fig. 2b separately compare measurements recorded from $X_b$, before and after the $N_2$-shift was executed. For $X_a$, which is initially detuned by $\Delta\lambda=+2~nm$ from the cavity, we observe clear triggered single photon generation with a multi-photon emission probability of $P_{X_a}=17.6$~\%. The spontaneous emission lifetime is $\tau_{X_a}=0.7\pm0.1~ns$ (Fig. 2a), slightly larger than the typical lifetime of dots in unpatterned bulk GaAs for this sample ($\tau_{bulk}=0.6~ns$) but much shorter than the typical lifetime of dots emitting deeply within the photonic bandgap ($4-12~ns$ \cite{Kaniber08}). The multi-photon emission probability of $X_b$ (Fig. 2b - upper panel) is clearly enhanced ($P_{X_b}=56.6$~\%) when compared to $X_a$, indicating that background emission from the cavity mode also contributes to the measured count rate within our detection bandwidth. This conclusion is not only supported by the observation that $X_b$ is superimposed on a broadband emission (gray shaded region in Fig. 1a), which is absent after N$_2$-deposition (Fig. 1b), but also by auto-correlation measurements of the same transition after N$_2$-deposition (Fig. 2b - lower panel). Due to the considerably longer lifetime $\tau_{X_b}^{N_2}=4.28\pm0.13~ns$ when the mode is detuned, the peaks in the auto-correlation spectrum broaden leading to a quasi cw-like result since $\tau_{X_b}^{N_2}\approx1/f_{laser}$. As for the in-resonance case, discussed above, we observe pronounced photon antibunching albeit with a smaller multi-photon emission probability of $P_{X_b}^{N_2}=25.0$~\% compared to the value of $P_{X_b}=56.6$~\%\ before the $N_2$-deposition. This clearly indicates that shifting the mode away from the dot enhances the purity of single photon emission.\\
Further insights into the dot-cavity system investigated and the nature of the background emission are obtained by examining the time-resolved data presented in Fig. 2b.  Detecting on $X_b$ we observe a biexponential decay transient before $N_2$-deposition but a clear monoexponential decay afterwards. We attribute the faster component of the biexponential decay $\tau_{X_b}=0.25\pm0.1~ns$ to the Purcell enhanced emission of the coupled $X_b$-transition and the slower lifetime ($\tau_{slow}=0.89\pm0.1~ns$) to the background emission into the cavity mode. Further support for this conclusion is provided by analyzing the lifetime-amplitude products of the fast and slow components of the decay, namely $\frac{A_{slow}\cdot\tau_{slow}}{A_{X_b}\cdot\tau_{X_b}}\sim23\%$. This value compares very well with the measured ratio of the intensities of $X_b$ and the mode background emission $\frac{PL_{Back.}}{PL_{X_B}}\sim22\%$ observed in $\mu$-PL (Fig. 1), confirming the identification of the fast and slow time constants.  Furthermore, by comparing $\tau_{X_b}$ measured before and after the $N_2$ shift we estimate the degree of spatial coupling of the QD to the cavity mode to be $|\frac{E(\vec{r})}{E_{max}}|\geq 40\%$ \cite{Englund05}, demonstrating that we deal with a dot-cavity system that is spectrally \textit{and} spatially well coupled. Since we spectrally select only one of the QD emission lines, namely $X_b$, this provides further support that the cavity mode background is responsible for the enhanced multi-photon emission probability observed in Fig. 2 before the $N_2$ shift.  \\
%
\begin{figure}[ht]
    \begin{center}
        \includegraphics[width=7cm]{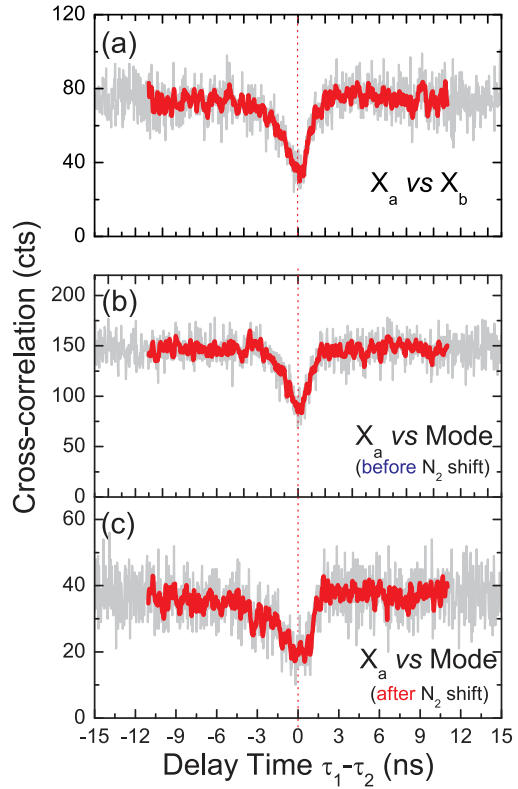}
    \end{center}
    \caption{ (color online) Cross-correlation measurements between (a) $X_a$ and $X_b$, (b) between $X_a$ and cavity mode
              \textit{before} N$_2$-deposition and (c) between $X_a$ and cavity mode \textit{after} N$_2$-deposition. We show the original data in gray 							 and the corresponding smoothed data in red.}
\end{figure}
%
%
To prove that the emission from $X_a$ and $X_b$ stem from the same quantum emitter and to further probe the nature of the cavity mode background, we performed cross-correlation measurements. In Fig. 3a we show the cw cross-correlation measurement between $X_a$ and $X_b$, before the $N_2$-shift was performed \cite{comment00}. A pronounced dip is observed at $\tau_1-\tau_2=0$, clearly showing that both transitions stem from the same QD, the different lifetimes of the two quantum states giving rise to the observed asymmetric character of the dip \cite{Kiraz02}. We also performed cross-correlation measurements between $X_a$ and the mode background emission, besides $X_b$ at $943.0$~nm (Fig. 3b). The clear dip in the cross-correlation histogram (Fig. 3b) shows unambiguously that the background emission from the mode arises from the same QD. Most remarkably, we measure a similar cross-correlation between $X_a$ and the cavity mode after N$_2$-deposition when the mode has been shifted by $+7.7~nm$ to longer wavelengths. Anticorrelations were found to persist as the detuning was systematically increased from $<2~nm$ up to $>10~nm$ \cite{comment01}. This implies the existence of an intrinsic dot-cavity coupling mechanism that is active for dot-cavity detunings up to several $10~nm$. Clearly, this phenomenon cannot be explained by the simplified atom-like picture that has been so successfully applied to quantum dots up to now.\\
%
%
\begin{figure}[ht]
    \begin{center}
   \includegraphics[width=6cm]{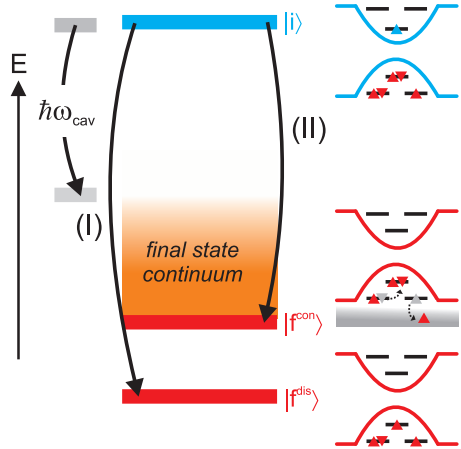}
    \end{center}
    \caption{ (color online) Energy diagram with initial state $|i>$ and two final state configurations $|f^{con}>$ and $|f^{dis}>$ and the
               corresponding QD occupations for a heavily positively charged dot.}
\end{figure}
%
%
Whilst the precise mechanism responsible for non-resonant dot-cavity coupling is not yet fully clear, we can already exclude a number of possibilities that have been discussed in the literature: Emission from 2D-2D wetting layer states, crossed (continuum) 0D-2D transitions \cite{Vasanelli02, Oulton03} and excited state QD transitions can all be excluded since they result in emission at higher energy than the fundamental exciton transition in the dot. Only mechanisms that lead to a \textit{reduction} in energy, and emission of a photon with \textit{lower} energy than the fundamental transition, could contribute to the emission of the cavity mode in our scenario. Recently, Press \emph{et al} \cite{Press07} suggested that the non-resonant coupling might be mediated by the emission of acoustic phonons. Phonon assisted processes are extremely unlikely for such large detunings, since the exciton-acoustic phonon coupling strength typically being strongest for phonon wavevectors $q_{ph}^{max}\sim\pi/d$, where $d$ is the characteristic confinement lengthscale in the dot. For $E_{ph}\sim10$~meV, the phonon wavevector is $q_{ph}=E_{ph}/{v_{s}\hbar} \sim 10 q_{ph}^{max}$ and we estimate that the acoustic phonon coupling strength is more than six orders of magnitude weaker as compared to the case for $q_{ph}\sim q_{ph}^{max}$ \cite{Krummheuer02}. We have found that the correlations between QD and cavity mode persist for energy detunings in excess of 19 meV (c.f. \cite{comment01}) and conclude, therefore, that acoustic phonon mediated processes cannot account for the observed behavior.\\
Instead, we suggest a mechanism depicted schematically in Fig. 4 as being responsible for the non-resonant dot-cavity coupling observed in our experiment. If indeed our dots are charged, possibly due to charge trapping at the etched GaAs surface in the PC, then it is feasible that we could encounter a scenario where charged exciton decay takes place into a \emph{continuum} of final states. This idea is illustrated in Fig. 4 for the case of a dot charged with $n$-holes. In the initial state of the transition, labeled $|i>$ in  Fig. 4, all particles are accommodated by the orbital states of the dot due to the balance between attractive and repulsive Coulomb interactions in the $X^{n^{+}}$ charged exciton initial state. However, in the $n$-hole final state of the transition, configurations exist where holes are distributed between the dot and the WL continuum, labeled by $|f^{con}>$ in Fig. 4. The absolute energy of this final state continuum would be determined by the interplay between the repulsive Coulomb interaction in the $n$-hole final state and the kinetic energy cost of ejecting particles into the continuum \cite{Ediger07}.  Exciton decay can then take place into discrete \emph{and} continuous final states, the branching ratio depending on the relative oscillator strength for each.  We suggest that decay into such a delocalized final state continuum may be enhanced close to the cavity energy giving rise to the observed behavior.  This process would explain the observed background emission in the cavity mode and also explain both the different decay lifetimes measured for the dot and cavity and the observation of correlations between the photons emitted ($X^{n+}$) state. We note that such a non-resonant coupling must occur \emph{during} photon emission, the mechanism being similar to the photo-induced hybridization of quantum levels recently observed for highly charged QDs \cite{Karrai04}.\\
%
In summary, we presented investigations of single photon generation and dot-cavity coupling effects for self-assembled QDs in PC defect nanocavities.  We have shown that, for dots that are spectrally and spatially coupled to the cavity, single photon generation is observed with elevated multi-photon emission probability. This was shown to be caused by background emission in the cavity mode. By shifting the cavity mode away from the dot by adsorption of molecular $N_2$ into the PC cavity we studied photon cross-correlations between the dot transitions and the detuned cavity mode. Pronounced cross-correlations were observed, that persist even for very large energy detunings ($>15~meV$). We argue that the coupling between QD and mode is mediated by photon mediated shake-up like processes, in which the final state of the transition is an excited continuum.\\
\\
We gratefully acknowledge financial support of the Deutsche Forschungsgemeindschaft via the Sonderforschungsbereich 631, Teilprojekt B3 and the German Excellence Initiative via the ``Nanosystems Initiative Munich (NIM)".\\


\begin{thebibliography}
\bibitem{comment01} Anticorrelations were observed for detunings up to 19 nm on a different QD - cavity system.
\end{thebibliography}
\end{document}